\documentclass[11pt,a4paper]{article}
\usepackage{amsfonts,amssymb}
\usepackage{amsopn}
\usepackage{amsmath,amsthm}
\usepackage{verbatim}

\DeclareMathOperator{\tr}{tr}

\DeclareMathAlphabet{\mathpzc}{OT1}{pzc}{m}{it}

\theoremstyle{plain}
\newtheorem{thm}{Theorem}

\newtheorem{lem}[thm]{Lemma}

\newtheorem{prop}[thm]{Proposition}

\theoremstyle{remark}
\newtheorem*{rem}{Remark}

\title{Some spectral bounds for Schr\"odinger operators with Hardy-type potentials}

\author{Douglas Lundholm\thanks{e-mail: dogge@math.kth.se}}

\date{\scriptsize{Department of Mathematics, Royal Institute of Technology\\SE-100 44 \ Stockholm, Sweden}}

\begin{document}

\maketitle

\begin{abstract}
	This note points
	out some bounds for the number of negative eigenvalues
	of Schr\"odinger operators with Hardy-type potentials,
	which follow from a simple coordinate transformation,
	and could prove useful in a spectral analysis of certain
	supersymmetric quantum mechanical models.
\end{abstract}

\vspace{1.5cm}

\setlength\arraycolsep{2pt}
\def\arraystretch{1.2}

\section{Introduction}

In a recent approach \cite{weighted-toy} to study the spectrum of a class of 
quantum mechanical models,
called supersymmetric matrix models and 
described by matrix-valued Schr\"odinger operators
(see e.g. \cite{Taylor,octonionic}),
it is relevant to consider the negative spectrum
of Schr\"odinger operators with critical Hardy terms, 
i.e. operators of the form
\begin{equation} \label{Hardy-operator}
	H = -\Delta_{\mathbb{R}^d} - \frac{(d-2)^2}{4|x|^2} + V(x),
\end{equation}
where $V$ is a real- or operator-valued potential.
This approach has so far only been applied to a simplified model,
where a bound for the number of negative eigenvalues of a one-dimensional
Schr\"odinger operator with Hardy term, following from a simple coordinate
transformation, turned out to be very important. 
The aim of this note is to extend this transformation to higher
dimensions and derive corresponding bounds which could be useful in an
extension of the technique to the higher-dimensional matrix models.
It also allows for
generalizations of some statements in \cite{Birman-Laptev,Weidl,Ekholm-Frank,Ekholm-Frank-halfline} 
regarding the one-dimensional, and higher-dimensional, operators. 
After searching 
the literature,
we found that the transformation we use and  
some of its consequences have been considered before 
(see e.g. \cite{Seto,Egorov-Kondratev} for the one-dimensional case, 
and \cite{Chadan_et_al} for higher dimensions), however, 
we are not aware of any reference stating these explicit bounds.
In Section 2 we recall some Hardy-type inequalities,
while the essential coordinate transformation is considered in Section 3,
and the bounds for the negative eigenvalues are stated and proved in Section 4.

\section{Some Hardy-type inequalities}

In the following we will denote by $\bar{B}_r(x)$ 
the closed ball of radius $r \ge 0$ at $x \in \mathbb{R}^d$.
We also use the conventions $\mathbb{R}_+ := (0,\infty)$ and $\mathbb{N} := \{0,1,2,\ldots\}$.

For $x \in \mathbb{R}^d \smallsetminus \{0\}$, $d=1,2,3,\ldots$, let
\begin{eqnarray*}
	\Psi_d(x) &:=& |x|^{-(d-2)},\quad \textrm{and} \\
	\Phi(x)   &:=& \ln |x|.
\end{eqnarray*}
These are the fundamental solutions of the Laplace operator
on $\mathbb{R}^{d \neq 2}$ and $\mathbb{R}^2$, respectively, 
since in the sense of distributions
\begin{eqnarray*}
	-\Delta_{\mathbb{R}^d}\Psi_d(x) &=& c_d \delta(x),\quad \textrm{and} \\
	-\Delta_{\mathbb{R}^2}\Phi(x)   &=& c_2 \delta(x),
\end{eqnarray*}
for some constants $c_d$ and $c_2$.
By considering the square root of these functions, 
we can prove the following Hardy-type inequalities.

\begin{prop} \label{prop_Hardy_type_ineq}
	We have
	\begin{equation} \label{normal_Hardy}
		-\Delta_{\mathbb{R}^d} - \frac{(d-2)^2}{4|x|^2} \quad \ge \quad 0,
	\end{equation}
	considered as a quadratic form on 
	$C_0^{\infty}(\mathbb{R}^d \smallsetminus \{0\})$,
	and
	\begin{equation} \label{extended_Hardy}
		-\Delta_{\mathbb{R}^d} - \frac{(d-2)^2}{4|x|^2} - \frac{1}{4|x|^2(\ln |x|)^2} \quad \ge \quad 0,
	\end{equation}
	considered as a quadratic form on 
	$C_0^{\infty}(\mathbb{R}^d \smallsetminus \bar{B}_{1}(0))$.
	In other words,
	\begin{equation} \label{normal_Hardy_ineq}
		\frac{(d-2)^2}{4} \int_{\mathbb{R}^d} \frac{|u|^2}{|x|^2} dx \quad \le \quad
		\int_{\mathbb{R}^d} |\nabla u|^2 dx,
	\end{equation}
	(the standard Hardy inequality in $L^2(\mathbb{R}^d)$)
	for all $u \in C_0^{\infty}(\mathbb{R}^d \smallsetminus \{0\})$, and
	\begin{equation} \label{extended_Hardy_ineq}
		\frac{(d-2)^2}{4} \int_{\mathbb{R}^d} \frac{|u|^2}{|x|^2} dx 
		+ \frac{1}{4} \int_{\mathbb{R}^d} \frac{|u|^2}{|x|^2(\ln |x|)^2} dx \quad \le \quad
		\int_{\mathbb{R}^d} |\nabla u|^2 dx,
	\end{equation}
	for all $u \in C_0^{\infty}(\mathbb{R}^d \smallsetminus \bar{B}_{1}(0))$.
\end{prop}

\begin{proof}
	Let us consider the first inequaliy \eqref{normal_Hardy}.
	It is straightforward to check that
	$$
		\nabla \ln \Psi_d(x)^{\frac{1}{2}} 
		= \frac{1}{2} \Psi_d(x)^{-1} (\nabla \Psi_d(x))
		= -\frac{d-2}{2} \frac{x}{|x|^2},
	$$
	and
	$$
		\Delta \ln \Psi_d(x)^{\frac{1}{2}}
		= -\frac{(d-2)^2}{2|x|^2}.
	$$
	Now, define the vector-valued operator
	$$
		Q := \nabla + \dot{\nabla} \ln \Psi_d(\dot{x})^{\frac{1}{2}}
		= \nabla + \frac{1}{2} \Psi_d(x)^{-1} \dot{\nabla} \Psi_d(\dot{x}).
	$$
	Then, considered in the sense of quadratic forms on 
	$C_0^{\infty}(\mathbb{R}^d \smallsetminus \{0\})$, we have
	\begin{eqnarray*}
		0 &\le& Q \cdot Q^\dagger 
		= (\nabla + \dot{\nabla} \ln \Psi_d(\dot{x})^{\frac{1}{2}}) \cdot (-\nabla + \dot{\nabla} \ln \Psi_d(\dot{x})^{\frac{1}{2}}) \\
		&=& -\nabla \cdot \nabla + \nabla \cdot \dot{\nabla} \ln \Psi_d(\dot{x})^{\frac{1}{2}} 
		- \dot{\nabla} \ln \Psi_d(\dot{x})^{\frac{1}{2}} \cdot \nabla + |\dot{\nabla} \ln \Psi_d(\dot{x})^{\frac{1}{2}}|^2 \\
		&=& -\Delta + \dot{\Delta} \ln \Psi_d(\dot{x})^{\frac{1}{2}} + |\dot{\nabla} \ln \Psi_d(\dot{x})^{\frac{1}{2}}|^2 \\
		&=& -\Delta - \frac{(d-2)^2}{2|x|^2} + \frac{(d-2)^2}{4|x|^2},
	\end{eqnarray*}
	which gives \eqref{normal_Hardy}.
	
	For the second inequality \eqref{extended_Hardy}, we observe that
	$$
		\nabla \ln \Phi(x)^{\frac{1}{2}} 
		= \frac{1}{2} \Phi(x)^{-1} (\nabla \Phi(x))
		= \frac{1}{2(\ln |x|)} \frac{x}{|x|^2},
	$$
	and
	$$
		\Delta \ln \Phi(x)^{\frac{1}{2}}
		= \nabla \cdot \frac{1}{2(\ln |x|)} \frac{x}{|x|^2}
		= -\frac{1}{2|x|^2(\ln |x|)^2} + \frac{d-2}{2|x|^2 \ln |x|}.
	$$
	Hence, defining
	$$
		\tilde{Q} := \nabla + \dot{\nabla} \ln (\Psi_d(\dot{x}) \Phi(\dot{x}))^{\frac{1}{2}}
		= \nabla + \frac{1}{2} \Psi_d(x)^{-1} \dot{\nabla} \Psi_d(\dot{x}) + \frac{1}{2} \Phi(x)^{-1} \dot{\nabla} \Phi(\dot{x}),
	$$
	we obtain, in the sense of quadratic forms on
	$C_0^{\infty}(\mathbb{R}^d \smallsetminus \bar{B}_1(0))$,
	\begin{eqnarray*}
		0 &\le& \tilde{Q} \cdot \tilde{Q}^\dagger \\
		&=& (\nabla + \dot{\nabla} \ln \Psi_d(\dot{x})^{\frac{1}{2}} + \dot{\nabla} \ln \Phi(\dot{x})^{\frac{1}{2}}) 
			\cdot (-\nabla + \dot{\nabla} \ln \Psi_d(\dot{x})^{\frac{1}{2}} + \dot{\nabla} \ln \Phi(\dot{x})^{\frac{1}{2}}) \\
		&=& -\Delta 
			+ \dot{\Delta} \ln \Psi_d(\dot{x})^{\frac{1}{2}} + \dot{\Delta} \ln \Phi(\dot{x})^{\frac{1}{2}} 
			+ 2 \dot{\nabla} \ln \Psi_d(\dot{x})^{\frac{1}{2}} \cdot \dot{\nabla} \ln \Phi(\dot{x})^{\frac{1}{2}} \\
		&&	+\ |\dot{\nabla} \ln \Psi_d(\dot{x})^{\frac{1}{2}}|^2 + |\dot{\nabla} \ln \Phi(\dot{x})^{\frac{1}{2}}|^2 \\
		&=& -\Delta - \frac{(d-2)^2}{2|x|^2} - \frac{1}{2|x|^2(\ln |x|)^2} + \frac{d-2}{2|x|^2 \ln |x|} - 2 \frac{d-2}{2|x|} \frac{1}{2|x| \ln |x|} \\
		&&	+\ \frac{(d-2)^2}{4|x|^2} + \frac{1}{4(\ln |x|)^2|x|^2},
	\end{eqnarray*}
	which proves \eqref{extended_Hardy}.
\end{proof}

\begin{rem}
	Note that if $\tilde{Q}$ is considered as taking values
	in the grade-one part of $\mathcal{G}(\mathbb{R}^d)$, the Clifford algebra over $\mathbb{R}^d$,
	(hence a Dirac-type operator)
	then also the Clifford product 
	$\tilde{Q} \tilde{Q}^\dagger = \tilde{Q} \cdot \tilde{Q}^\dagger + \tilde{Q} \wedge \tilde{Q}^\dagger$
	(decomposed in terms of inner and outer products)
	is a non-negative operator on e.g.
	$C_0^\infty(\mathbb{R}^d \smallsetminus \bar{B}_1(0)) \otimes \mathcal{S}$,
	where $\mathcal{S}$ denotes a representation of $\mathcal{G}(\mathbb{R}^d)$.
\end{rem}

\section{Transformation of quadratic forms}

Combining the so-called ground state representation of 
the operator \eqref{Hardy-operator} 
(which is implicitly used in Proposition \ref{prop_Hardy_type_ineq}),
with a coordinate transformation,
we can relate a Schr\"odinger operator with a Hardy term
defined on a domain in $\mathbb{R}^d$
to a corresponding operator without the term on a transformed domain.
More precisely, 
denoting $B_R^c := \mathbb{R}^d \smallsetminus \bar{B}_R(0)$, for $R \ge 0$,
and for $R < 0$ generalizing this to denote 
the cone parametrized by $(r,\omega) \in (R,\infty) \times S^{d-1}$
(for $d=1$ we write $B_R^c := (R,\infty)$),
we have the following simple result.

\begin{lem} \label{lem_transf_Schroedinger}
	For any $u \in C_0^\infty(B_R^c)$, 
	$R \ge 0$ for $d$ odd, $R \ge 1$ for $d$ even,
	we have
	\begin{eqnarray*} 
		\lefteqn{ 
		\left\langle u, \left( -\Delta_{\mathbb{R}^d} - \frac{(d-2)^2}{4|x|^2} + V(x) \right) u \right\rangle_{L^2(B_R^c)} } \\
		&=& \left\langle \psi, \left( -\Delta_{\mathbb{R}^d} - \left(1 - \frac{1}{|x|^2}\right)\Delta_{S^{d-1}} - \frac{(d-1)(d-3)}{4|x|^2} + e^{2r}V(e^{r}\omega) \right) \psi \right\rangle_{L^2(B_{\ln R}^c)}
	\end{eqnarray*}
	where $\psi(r\omega) := r^{-\frac{d-1}{2}} e^{\frac{d-2}{2} r} u(e^r \omega)$,
	and $(r,\omega) \in (\ln R, \infty) \times S^{d-1}$. 
\end{lem}

\begin{proof}
	In spherical coordinates, the l.h.s., denote it $I$, is 
	$$
			\int_R^\infty \int_{\omega \in S^{d-1}} \overline{u(r,\omega)} \left( 
				-\frac{1}{r^{d-1}} \partial_r r^{d-1} \partial_r - \frac{1}{r^2}\Delta_{\omega} - \frac{(d-2)^2}{4r^2} + V(r\omega)
			\right) u(r,\omega) r^{d-1} dr d\omega.
	$$
	First, put $u(x) =: \Psi_d(x)^{\frac{1}{2}} v(x)$, 
	i.e $v(r,\omega) := r^{\frac{d-2}{2}}u(r,\omega)$, 
	and arrive via partial integration at the ground state representation,
	\begin{eqnarray*}
	I	&=& \int_R^\infty \int_{\omega \in S^{d-1}} \left( 
				|\partial_r u(r,\omega)|^2 - \frac{(d-2)^2}{4r^2} |u|^2 + \overline{u} \left(-\frac{1}{r^2}\Delta_{\omega} + V(r\omega)\right) u 
			\right) r^{d-1} dr d\omega \\
		&=& \int_R^\infty \int_{S^{d-1}} \left( 
				|\partial_r v(r,\omega)|^2 + \overline{v} \left(-\frac{1}{r^2}\Delta_{\omega} + V(r\omega)\right) v 
			\right) r dr d\omega.
	\end{eqnarray*}
	Because of the form of the integral measure here, 
	this expression actually possesses two-dimensional features,
	which explains 
	why the function $\Phi$ enters in the proof of Proposition \ref{prop_Hardy_type_ineq}.
	Next, change variables, $r =: e^s$, $dr = rds$, $w(s,\omega) := v(e^s,\omega)$,
	which in a sense lowers the dimension by one:
	\begin{eqnarray}
	I	&=& \int_{\ln R}^\infty \int_{S^{d-1}} \left( 
				|\partial_s w(s,\omega)|^2 + \overline{w} \left( -\Delta_{\omega} + e^{2s}V(e^s\omega) \right) w 
			\right) ds d\omega \nonumber \\
		&=& \left\langle w, \left( 
				-\partial_s^2 - \Delta_{S^{d-1}} + e^{2s}V(e^s\omega) 
			\right) w\right\rangle_{L^2((\ln R,\infty)) \otimes L^2(S^{d-1})}. \label{one-dim_groundstate_rep}
	\end{eqnarray}
	Finally, transform back from this corresponding ground state representation,
	by taking $\psi(s,\omega) := s^{-\frac{d-1}{2}} w(s,\omega)$, resulting in
	\begin{eqnarray*}
	I	&=& \int_{\ln R}^\infty \int_{S^{d-1}} \overline{\psi} \left( 
				-\frac{1}{s^{d-1}} \partial_s s^{d-1} \partial_s - \frac{(d-1)(d-3)}{4s^2} - \Delta_{\omega} + e^{2s}V(e^s\omega) 
			\right) \psi \ s^{d-1} ds d\omega \\
		&=& \left\langle \psi, \left( 
				-\Delta_{\mathbb{R}^d} + \left( \frac{1}{s^2} - 1 \right) \Delta_{\omega} - \frac{(d-1)(d-3)}{4s^2} + e^{2s}V(e^s\omega) 
			\right) \psi \right\rangle_{L^2(B_{\ln R}^c)},
	\end{eqnarray*}
	which is the r.h.s. of the claimed identity.
\end{proof}

In particular, we have the following special cases and consequences.

\begin{prop} \label{prop_transf_1d}
	Consider $d=1$.
	For all $u \in C_0^\infty(\mathbb{R}_+)$, we have
	$$
		\left\langle u, \left( -\frac{d^2}{dx^2} - \frac{1}{4x^2} + V(x) \right) u \right\rangle_{L^2(\mathbb{R}_+)}
		= \left\langle \psi, \left( -\frac{d^2}{dx^2} + e^{2x}V(e^{x}) \right) \psi \right\rangle_{L^2(\mathbb{R})}.
	$$
	Furthermore, if $V(x) = -\frac{1}{4x^2(\ln x)^2} + W(x)$ then 
	we have for all $u \in C_0^\infty((1,\infty))$
	\begin{eqnarray*}
		\lefteqn{ 
		\left\langle u, \left( -\frac{d^2}{dx^2} - \frac{1}{4x^2} - \frac{1}{4x^2(\ln x)^2} + W(x) \right) u \right\rangle_{L^2((1,\infty))} } \\
		&& \quad = \left\langle \phi, \left( -\frac{d^2}{dx^2} + e^{2x}e^{2e^x}W(e^{e^x}) \right) \phi \right\rangle_{L^2(\mathbb{R})},
	\end{eqnarray*}
	with $\phi(x) = e^{-x/2}\psi(e^x) = e^{-x/2}e^{-e^x/2}u(e^{e^x})$.
	This procedure can be iterated further to the interval $(e,\infty)$, and so on.
\end{prop}

\begin{prop} \label{prop_transf_2d}
	Consider $d=2$, with polar coordinates $(r,\varphi)$.
	For all $u \in C_0^\infty(B_1^c)$, we have
	\begin{eqnarray*}
		\lefteqn{ 
		\left\langle u, \left( -\Delta_{\mathbb{R}^2} + V(x) \right) u \right\rangle_{L^2(B_1^c)} } \\
		&& \quad = \left\langle \psi, \left( -\Delta_{\mathbb{R}^2} -(1 - r^{-2})\frac{d^2}{d\varphi^2} + \frac{1}{4r^2} + e^{2r}V(e^{r},\varphi) \right) \psi \right\rangle_{L^2(\mathbb{R}^2)},
	\end{eqnarray*}
	so that, with $V(x) = -\frac{1}{4|x|^2(\ln |x|)^2} + W(x)$
	we have for all $u \in C_0^\infty(B_1^c)$
	\begin{eqnarray*}
		\lefteqn{ 
		\left\langle u, \left( -\Delta_{\mathbb{R}^2} - \frac{1}{4r^2(\ln r)^2} + W(x) \right) u \right\rangle_{L^2(B_1^c)} } \\
		&=& \left\langle \psi, \left( -\Delta_{\mathbb{R}^2} - (1 - r^{-2})\frac{d^2}{d\varphi^2} + e^{2r}W(e^r,\varphi) \right) \psi \right\rangle_{L^2(\mathbb{R}^2)}.
	\end{eqnarray*}
\end{prop}

\begin{prop} \label{prop_transf_any-d}
	For general $d=1,2,3,\ldots$, we have
	\begin{eqnarray*}
		\lefteqn{ 
		\left\langle u, \left( -\Delta_{\mathbb{R}^d} - \frac{(d-2)^2}{4|x|^2} - \frac{1}{4|x|^2(\ln |x|)^2} + V(x) \right) u \right\rangle_{L^2(B_1^c)} } \\
		&=& \left\langle \psi, \left( -\Delta_{\mathbb{R}^d} - \left(1 - \frac{1}{|x|^2}\right)\Delta_{S^{d-1}} - \frac{(d-2)^2}{4|x|^2} + e^{2r}V(e^{r}\omega) \right) \psi \right\rangle_{L^2(\mathbb{R}^d)}
	\end{eqnarray*}
	for all $u \in C_0^\infty(B_1^c)$,
	where $\psi(r\omega) = r^{-\frac{d-1}{2}} e^{\frac{d-2}{2} r} u(e^r \omega)$.
\end{prop}
\begin{proof}
	This follows immediately from Lemma \ref{lem_transf_Schroedinger} because
	$(d-1)(d-3) + 1 = (d-2)^2$.
\end{proof}

\begin{rem}
	The above transformations all extend to the case when $V$ is 
	operator-valued (cp. \cite{Hundertmark}).
\end{rem}

\noindent
In the following, denote
$$
	\ln^{(n)} x := \underbrace{\ln \circ \ln \circ \ldots \circ \ln}_{\textrm{$n$ factors}} (x) 
	\quad \textrm{and} \quad
	\exp^{(n)} x := \underbrace{\exp \circ \exp \circ \ldots \circ \exp}_{\textrm{$n$ factors}} (x).
$$
Then we also obtain by iteration of Lemma \ref{lem_transf_Schroedinger}
the following generalization of Proposition \ref{prop_Hardy_type_ineq}:

\begin{prop} \label{prop_general_Hardy}
	For general $d=1,2,3,\ldots$, we have
	\begin{eqnarray*}
		-\Delta_{\mathbb{R}^d} - \frac{(d-2)^2}{4|x|^2} - \frac{1}{4|x|^2(\ln |x|)^2} - \ldots - \frac{1}{4|x|^2(\ln |x|)^2 \ldots (\ln^{(n)} |x|)^2} \quad \ge 0
	\end{eqnarray*}
	in the sense of quadratic forms on 
	$C_0^\infty \left( B_{\exp^{(n)} 0}^c \right)$.
\end{prop}

\section{Bounds for the number of negative eigenvalues}

Denote by $N(A)$ the rank of the spectral projection on $(-\infty,0)$
of a self-adjoint operator $A$,
and by $V_{\pm}$ the positive/negative parts of a function $V$.
In the one-dimensional case we have the following 
(cp. e.g. Proposition 3.2 in \cite{Ekholm-Frank} and Theorem 9 in Chapter 8 of \cite{Egorov-Kondratev}):

\begin{thm} \label{CLR-bound_one-dim}
	Let $n \in \mathbb{N}$, and $V$ be a real-valued potential such that \\ 
	$x^2(\ln x)^2 \ldots (\ln^{(n)} x)^2 V(x)$
	is bounded from below. 
	Then the self-adjoint operator
	$$
		H_{1,n} := -\frac{d^2}{dx^2} - \frac{1}{4x^2} - \ldots - \frac{1}{4x^2(\ln x)^2 \ldots (\ln^{(n)} x)^2} + V(x),
	$$
	defined by Friedrichs extension on $C_0^\infty \left( (\exp^{(n)} 0, \infty) \right)$, 
	has at least one negative eigenvalue for all negative (non-zero) potentials $V$.
	Furthermore, the number of negative eigenvalues is bounded by
	$$
		N(H_{1,n}) \le 1 + \int_{\exp^{(n)} 0}^\infty |V(x)_-| |x| |\ln x| \ldots |\ln^{(n+1)} x| \thinspace dx.
	$$
	On the other hand, $H_{1,n}$ defined by Friedrichs extension 
	on $C_0^\infty \left( (\exp^{(n)} 1, \infty) \right)$ satisfies the bound
	$$
		N(H_{1,n}) \le \int_{\exp^{(n)} 1}^\infty |V(x)_-| |x| |\ln x| \ldots |\ln^{(n+1)} x| \thinspace dx.
	$$
\end{thm}

\begin{rem}
	For a different version of the latter bound (for $n=0$)
	in the case of operator-valued potentials, and an application,
	see \cite{weighted-toy}.
\end{rem}

\begin{proof}
	We will use that
	Bargmann's bound in three dimensions,
	together with Dirichlet boundary conditions, implies
	(see e.g. \cite{Reed-Simon})
	$$
		N\left( -\frac{d^2}{dx^2}\big|_\mathbb{R} + V(x) \right) \le 1 + \int_{-\infty}^\infty |V(x)_-| |x| \thinspace dx.
	$$
	By Proposition \ref{prop_transf_1d}, 
	we have for any $u \in C_0^\infty((\exp^{(n)} 0, \infty))$
	\begin{equation} \label{one-dim_quad_form_relation}
		\langle u, H_{1,n} u \rangle_{L^2((\exp^{(n)} 0,\infty))} 
		= \left\langle \phi, \left( 
			-\frac{d^2}{dx^2}\big|_\mathbb{R} + e^{2x} \ldots e^{2\exp^{(n)}x} V(\exp^{(n+1)}x) 
			\right) \phi \right\rangle_{L^2(\mathbb{R})},
	\end{equation}
	for some $\phi \in C_0^\infty(\mathbb{R})$. 
	From this expression one immediately obtains the first
	statement of the theorem by relating to the case for 
	a one-dimensional Schr\"odinger operator.
	Furthermore, linearly independent sets of such functions $u$
	correspond to linearly independent sets of $\phi$.
	Hence, since $N(H_{1,n})$ is equal to the maximal dimension of a subspace
	of functions $u \in C_0^\infty((\exp^{(n)} 0, \infty))$ 
	s.t. $\langle u,H_{1,n}u \rangle < 0$,
	and correspondingly for the operator on the r.h.s. of \eqref{one-dim_quad_form_relation},
	we have
	\begin{eqnarray*}
		N(H_{1,n}) &=& N\left( -\frac{d^2}{dx^2}\big|_\mathbb{R} + e^{2x} \ldots e^{2\exp^{(n)} x} V(\exp^{(n+1)}x) \right) \\
		&\le& 1 + \int_{-\infty}^{\infty} e^{2x} \ldots e^{2\exp^{(n)}x} |V(\exp^{(n+1)}x)_-| |x| \thinspace dx \\
		&=& 1 + \int_{\exp^{(n)} 0}^\infty |V(y)_-| |y| |\ln y| \ldots |\ln^{(n+1)} y| \thinspace dy.
	\end{eqnarray*}
	The second bound is proved analogously, using that
	\begin{equation*}
		N\left( -\frac{d^2}{dx^2}\big|_{\mathbb{R}_+} + V(x) \right) \le \int_{0}^\infty |V(x)_-| |x| \thinspace dx. \qedhere
	\end{equation*}
\end{proof}

For higher dimensions we have instead the following version of 
the above bounds:

\begin{thm} \label{CLR-bound_higher-dim}
	Let $n \in \mathbb{N}$, $V$ be real-valued and s.t.
	$|x|^2(\ln |x|)^2 \ldots (\ln^{(n)} |x|)^2 V(x)$
	is bounded from below, and let
	$$
		H_{d,n} := -\Delta_{\mathbb{R}^d} - \frac{(d-2)^2}{4|x|^2} - \ldots - \frac{1}{4|x|^2(\ln |x|)^2 \ldots (\ln^{(n)} |x|)^2} + V(x)
	$$
	be defined as a self-adjoint operator by Friedrichs extension 
	on $C_0^\infty(B_{\exp^{(n+2)} 0}^c)$.
	For $d \ge 3$, and some universal positive constant $C_d$, we have the following 
	bound for the number of negative eigenvalues:
	\begin{eqnarray*}
		N(H_{d,n}) &\le& C_d \int\limits_{|x| > \exp^{(n+2)} 0} 
		\left( \frac{(d-1)(d-3)}{4|x|^2(\ln |x|)^2 \ldots (\ln^{(n+1)} |x|)^2} -  V(x) \right)_+^{\frac{d}{2}} \\
		&& \qquad \qquad \cdot
		(\ln |x|)^{d-1} \ldots (\ln^{(n+1)} |x|)^{d-1} \thinspace dx.
	\end{eqnarray*}
	On the other hand, $H_{d,n}$
	defined by Friedrichs extension on $C_0^\infty(B_{\exp^{(n+2)} 1}^c)$
	satisfies the bound
	\begin{eqnarray*}
		N(H_{d,n}) &\le& C_d \int\limits_{|x| > \exp^{(n+2)} 1} 
		\left( \frac{ (d-1)(d-3) - (\ln^{(n+2)} |x|)^2 }{4|x|^2(\ln |x|)^2 \ldots (\ln^{(n+2)} |x|)^2} -  V(x) \right)_+^{\frac{d}{2}} \\
		&& \qquad \qquad \cdot
		(\ln |x|)^{d-1} \ldots (\ln^{(n+2)} |x|)^{d-1} \thinspace dx.
	\end{eqnarray*}
\end{thm}

\begin{rem}
	These bounds also extend to operator-valued potentials 
	according to \cite{Hundertmark},
	where $(\ )_+^{\frac{d}{2}}$ is replaced by $\tr \thinspace (\ )_+^{\frac{d}{2}}$,
	and $C_d$ is slightly larger.
	Also, by a monotonicity argument (see e.g. Remark 2.2 in \cite{Hundertmark}),
	they imply corresponding Lieb-Thirring inequalities for non-zero moments of the eigenvalues 
	(cp. \cite{Ekholm-Frank}).
\end{rem}

\begin{rem}
	Note that there is always an extra contribution to the above bound
	for the number of negative eigenvalues of $H_{d,n}$ for all $d \ge 4$,
	but not so in the case of $d=1$ (Theorem \ref{CLR-bound_one-dim}) and $d=3$. 
	This is quite interesting when related with the fact
	that supersymmetric matrix models, 
	split into coordinates of $\mathbb{R}^d \times \mathbb{R}^2$ (cp. \cite{octonionic}),
	are conjectured to have zero energy states for $d=7$, but not for $d=0,1,3$.
\end{rem}

\begin{proof}
	Here we apply the Cwikel-Lieb-Rozenblum bound for $d \ge 3$ 
	(see e.g. \cite{Reed-Simon,Egorov-Kondratev}):
	$$
		N\left( -\Delta_{\mathbb{R}^d} + V(x) \right) \le C_d \int_{\mathbb{R}^d} |V(x)_-|^{\frac{d}{2}} dx.
	$$
	By iterating the bound obtained from Lemma \ref{lem_transf_Schroedinger},
	\begin{eqnarray*}
		\lefteqn{ 
		\left\langle u, \left( -\Delta_{\mathbb{R}^d} - \frac{(d-2)^2}{4|x|^2} + V(x) \right) u \right\rangle_{L^2(B_e^c)} } \\
		&\ge& \left\langle \psi, \left( -\Delta_{\mathbb{R}^d} - \frac{(d-1)(d-3)}{4|x|^2} + e^{2|x|}V(e^{|x|}\omega) \right) \psi \right\rangle_{L^2(B_1^c)},
	\end{eqnarray*}
	with $\psi(r\omega) = r^{-\frac{d-1}{2}} e^{\frac{d-2}{2} r} u(e^r \omega)$,
	we have as in the one-dimensional case
	\begin{eqnarray*}
		N(H_{d,n}) &\le& N\left( -\Delta_{B_1^c} - \frac{(d-1)(d-3)}{4|x|^2} + e^{2|x|} \ldots e^{2 \exp^{(n)} |x|} V\left( (\exp^{(n+1)} |x|)\omega \right) \right) \\
		&\le& C_d \int_{B_1^c} 
		\left( \frac{(d-1)(d-3)}{4|x|^2} - e^{2|x|} \ldots e^{2 \exp^{(n)} |x|} V\left( (\exp^{(n+1)} |x|)\omega \right) \right)_+^{\frac{d}{2}} dx \\
		&=& C_d \int_{B_{\exp^{(n+2)} 0}^c} 
		\left( \frac{(d-1)(d-3)}{4|x|^2(\ln |x|)^2 \ldots (\ln^{(n+1)} |x|)^2} -  V(x) \right)_+^{\frac{d}{2}} \\
		&& \qquad \qquad \cdot (\ln |x|)^{d-1} \ldots (\ln^{(n+1)} |x|)^{d-1} \thinspace dx.
	\end{eqnarray*}
	For the operator on the domain $B_{\exp^{n+2} 1}^c$,
	we can add and subtract a term $1/(4|x|^2)$ 
	and iterate one step further to obtain
	\begin{eqnarray*}
		N(H_{d,n})
		&\le& C_d \int_{B_1^c} 
		\left( \frac{(d-1)(d-3)}{4|x|^2} - \frac{1}{4} - e^{2|x|} \ldots e^{2 \exp^{(n+1)} |x|} V\left( (\exp^{(n+2)} |x|)\omega \right) \right)_+^{\frac{d}{2}} dx. 
	\end{eqnarray*}
	The stated bound then follows as above.
\end{proof}

We expect that it is possible to find analogous bounds on the larger domains
$B_{\exp^{(n)} 0}^c$ and $B_{\exp^{(n)} 1}^c$. 
Indeed, for central potentials we have 
the following:

\begin{thm}
	If $V(x) = \tilde{V}(|x|)$ is a central potential s.t. \\
	$r^2(\ln r)^2 \ldots (\ln^{(n)} r)^2 \tilde{V}(r)$
	is bounded from below, then for $H_{d,n}$ defined on the domain $B_{\exp^{(n)} 0}^c$
	\begin{eqnarray*}
		N(H_{d,n}) &\le& \sum_{l=0}^{l_\textup{max}} D_{d,l} \left( 
			1 + \int_{\exp^{(n)} 0}^\infty \left( -\frac{l(l+d-2)}{r^2} - \tilde{V}(r) \right)_+ |r| |\ln r| \ldots |\ln^{(n+1)} r| \thinspace dr
		\right),
	\end{eqnarray*}
	while on $B_{\exp^{(n)} 1}^c$
	\begin{eqnarray*}
		N(H_{d,n}) &\le& \sum_{l=0}^{l_\textup{max}} D_{d,l} 
			\int_{\exp^{(n)} 1}^\infty \left( -\frac{l(l+d-2)}{r^2} - \tilde{V}(r) \right)_+ r |\ln r| \ldots |\ln^{(n+1)} r| \thinspace dr,
	\end{eqnarray*}
	where
	$$
		D_{d,l} := \frac{(2l+d-2)\Gamma(d+l-2)}{\Gamma(d-1)\Gamma(l+1)},
	$$
	and $l_\textup{max}$ is the maximal integer $l \ge 0$ s.t. the negative part of
	$\frac{l(l+d-2)}{r^2} + \tilde{V}(r)$ is non-zero on the respective domain.
\end{thm}

\begin{proof}
	For central potentials, we can split the Hilbert space 
	$\mathcal{H} = \bigoplus_{l=0}^{\infty} \mathcal{H}_l$
	into eigenspaces of the the angular laplacian,
	where
	$
		-\Delta_{S^{d-1}}|_{\mathcal{H}_l} = l(l+d-2)
	$
	with degeneracy $D_{d,l}$ (see e.g. \cite{Vilenkin}; cp. \cite{Seto}).
	Using \eqref{one-dim_groundstate_rep} and iterating, we have for 
	$u = \tilde{u} \otimes \psi \in 
		C_0^\infty((\exp^{(n)} 0,\infty)) \otimes L^2(S^{d-1}) \cap \mathcal{H}_l$
	\begin{eqnarray*}
		\lefteqn{ \langle u, H_{d,n} u \rangle_{L^2(\mathbb{R}^d)} }\\
		&=& \left\langle w, \left( 
				-\partial_s^2 + e^{2s} \ldots e^{2\exp^{(n)} s} \left( \frac{l(l+d-2)}{(\exp^{(n+1)} s)^2} + \tilde{V}(\exp^{(n+1)} s) \right)
			\right) w\right\rangle_{L^2(\mathbb{R})} \\
		&& \quad \cdot \|\psi\|_{L^2(S^{d-1})}^2,
	\end{eqnarray*}
	with $w \in C_0^\infty(\mathbb{R})$.
	Hence, 
	by reasoning as in the proof of Theorem \ref{CLR-bound_one-dim},
	\begin{eqnarray*}
		N(H_{d,n}|_{\mathcal{H}_l})
		&\le& D_{d,l} 
		\left( 1 + \int_{-\infty}^\infty e^{2s} \ldots e^{2 \exp^{(n)} s} 
			\left( -\frac{l(l+d-2)}{(\exp^{(n+1)} s)^2} - \tilde{V}(\exp^{(n+1)} s) \right)_+ |s| \thinspace ds
			\right)
	\end{eqnarray*}
	if $l \le l_\textup{max}$, and $N(H_{d,n}|_{\mathcal{H}_l}) = 0$ otherwise.
	The first statement of the theorem then follows by a change of variables,
	and similar reasoning gives the second statement.
\end{proof}

\subsubsection*{Acknowledgements}

I am most grateful to Oleg Safronov for many useful discussions 
and for pointing out to me the
one-dimensional case (Proposition \ref{prop_transf_1d}
and a version of Theorem \ref{CLR-bound_one-dim}).
I would also like to thank Jens Hoppe for bringing us together,
as well as Ari Laptev for useful discussions.
This work was supported by the Swedish Research Council, 
the Knut and Alice Wallenberg Foundation (grant KAW 2005.0098),
and the European Science Foundation activity MISGAM.

\end{document}